%% file: main.tex
\documentclass[10pt,letterpaper,twocolumn]{article}

\usepackage[letterpaper,margin=0.75in]{geometry}
\usepackage[hyphens]{url}
\usepackage{graphicx}
\usepackage{natbib}
\usepackage{caption}
\usepackage{amsmath}
\usepackage{amssymb}
\usepackage{booktabs}
\usepackage{times}
\usepackage{microtype}
\usepackage[colorlinks=true,allcolors=blue]{hyperref}

\hypersetup{
  pdftitle={Auditable Release Control for Pedagogical Leakage in LLM Tutors},
  pdfauthor={Nizam Kadir},
  pdfsubject={Preprint; not peer reviewed}
}

\title{Auditable Release Control for Pedagogical Leakage in LLM Tutors}
\author{Nizam Kadir\\
Science, Mathematics and Technology\\
Singapore University of Technology and Design (SUTD)\\
Singapore\\
\texttt{nizam\_kadir@mymail.sutd.edu.sg}}
\date{\textit{Preprint. Not peer reviewed.}}

\begin{document}

\maketitle

\begin{abstract}
Large language model tutors can be correct and helpful yet disclose an answer
or decisive reasoning before that disclosure is authorized.  We formalize this
state- and action-dependent failure as \emph{pedagogical leakage} and introduce
an authorization-aware complete-mediation boundary.  A selector emits one of
five disclosure contracts, trusted policy gates privileged modes, and a
renderer proposes language.  A single release function applies inspectable
checks, optional cumulative verification, and action-specific fallback;
replayable traces separate selection, generation, verification, and enforcement
failures.

Matched component attribution exposes a safety--utility frontier.  On 599 fixed
Gemini 3.5 proposals, strict mediation reduces blinded three-model
panel-majority leakage flags from 181 to 0 (paired problem-cluster difference
$-30.22$ points, 95\% CI $[-35.00,-25.72]$), while replacing 581 responses and
lowering helpfulness.  Checker-triggered fallback alone yields 11 majority
flags; adding the semantic verifier yields 14 and no reliable marginal gain.  A
global $A_1$ scaffold yields 0 majority and 54 any-judge flags, outperforming
fitted Q on automatic safety and utility.  In an externally timestamped
replication over 40 unseen problem clusters and 480 attack sequences,
high-assurance release reduces majority flags from 42 to 8 ($-7.08$ points,
95\% CI $[-13.13,-2.29]$); seven failures persist, one is introduced, and mean
helpfulness falls by $.192$.  These results establish an auditable release
boundary and failure attribution under declared contracts, not universal
semantic safety or learning gains.
\end{abstract}

\section{Introduction}
An LLM tutor's most consequential output decision is not only what is correct,
but what may be released \emph{now}.  Intelligent tutoring systems traditionally
separate student modeling from the selection of problems, hints, and feedback
\cite{corbett1995knowledge,feng2009addressing}.  LLM tutors can collapse that
policy and its linguistic realization into one fluent generator
\cite{kasneci2023chatgpt,tack2022ai,macina2023mathdial,learnlm2024}, hiding when
a supportive response has silently completed the learner's reasoning.

We define \textbf{pedagogical leakage} as unauthorized answer disclosure,
decisive calculation disclosure, or over-constraining the search so strongly
that the learner no longer constructs the key step.  The same text can violate
an elicitation contract and be appropriate in an authorized worked solution.
This is unlike generic harmful-content moderation: the protected information is
problem-specific, its release depends on learner state and pedagogical action,
and authorization can legitimately change within a dialogue.  ICAP links
constructive engagement with learning \cite{chi2014icap}; scaffolding and
productive-failure research likewise treats bounded difficulty as useful
\cite{wood1976role,kapur2008productive}.

Preference alignment optimizes judgments over completed responses \cite{christiano2017deep,stiennon2020learning,rlhf_foundation,bai2022helpful}, but immediate helpfulness does not specify when a tutor should elicit, scaffold, explain, or release an answer. Recent pedagogical alignment, steering, and tutoring-policy methods address this conflict through preference optimization, reinforcement learning, or explicit planning \cite{sonkar-etal-2024-pedagogical,puech-etal-2025-towards,dinucu-jianu-etal-2025-problem,scarlatos2025training,li-etal-2026-planning}. Recent adversarial evaluations also show that answer-inducing prompts remain a distinct attack surface \cite{zhao-etal-2026-evaluating-answer,zhao-etal-2026-shape}. Our engineering question is complementary and narrower: can disclosure policy, language generation, authorization, and enforcement be separated so that failures are measurable and rejected releases cannot bypass the same decision point?

We introduce a modular \textbf{auditable release-control architecture}. A \emph{selector} emits a small pedagogical action; an authorization gate controls privileged disclosure modes; and a \emph{renderer} produces text. A two-mode guard applies a deterministic checker alone in permissive mode or adds a cumulative semantic verifier in high-assurance mode; either path substitutes an action-specific fallback on failure. This architecture does not make a universal semantic-safety guarantee. It makes the release path observable and permits matched ablations of the selector, renderer, and guard.

Our contributions are:
\begin{enumerate}
    \item We formalize pedagogical leakage as bounded, authorization-dependent disclosure and implement one fail-closed release function with a replayable trace.
    \item Matched component attribution shows that fallback enforcement causes the large automatic safety reduction and its utility cost; global $A_1$ beats fitted Q, and the semantic verifier adds no reliable marginal benefit in the matched lane.
    \item An externally timestamped prospective study quantifies residual failures under unseen problem clusters and adaptive attacks using paired cluster inference and per-model outcomes rather than treating model-panel votes as ground truth.
\end{enumerate}

\section{Related Work}
\textbf{Student modeling and tutoring policy.} Bayesian Knowledge Tracing (BKT) estimates latent mastery for learner--skill pairs \cite{corbett1995knowledge}; ASSISTments connected such models with large-scale tutoring logs \cite{feng2009addressing}. Educational data mining broadened interaction-log prediction \cite{baker2009state}, and offline policy work studies pedagogical decisions without online exploration \cite{mandel2014offline,gao2024headstart}. We use this literature to construct an auditable diagnostic state/action interface, not to infer treatment effects from heterogeneous event logs.

\textbf{Pedagogical alignment and tutor evaluation.} Benchmarks distinguish tutoring from answer production through dialogue pedagogy, mathematical scaffolding, and safety rubrics \cite{tack2022ai,macina2023mathdial,macina2025mathtutorbench,maurya2025unifying,hazra2026safetutors,scale2025tutorbench}. Evidence from deployed courses also shows that students actively extract answers, motivating turn-level computational measures alongside outcome studies \cite{kobler-etal-2026-students}. Sonkar et al. construct synthetic preferences for pedagogical alignment \cite{sonkar-etal-2024-pedagogical}; Puech et al. optimize prompts to follow a multi-turn productive-failure graph \cite{puech-etal-2025-towards}; recent policy work uses online RL or multi-horizon preference optimization \cite{dinucu-jianu-etal-2025-problem,shi-etal-2026-beyond}; and Scarlatos et al. optimize candidate tutor utterances for predicted student correctness and pedagogy \cite{scarlatos2025training}. ScaffoldLM explicitly plans and tracks dialogue progress \cite{li-etal-2026-planning}. These methods optimize tutoring behavior. We instead study whether a separately selected disclosure contract is mediated at release time. This distinction matters because simply prompted simulated students can be behaviorally unreliable \cite{scarlatos-etal-2026-simulated}.

The closest AAAI-26 systems sharpen this distinction. EduGuardBench evaluates
pedagogical fidelity and adversarial safety across 14 models after calibrating an
LLM judge against 200 human labels \cite{jiang2026eduguardbench}; it measures
model behavior, whereas our system enforces and attributes a release decision.
LEAP plans long-horizon scaffolds from cognitive state \cite{dong2026leap}; our
boundary operates after action selection and claims no learner benefit.
VerifyBench uses expert annotation to expose verifier sensitivity to response
form and domain \cite{li2026verifybench}, motivating independent calibration of
our narrower authorization-dependent verifier.

\textbf{Adversarial answer release.} Zhao et al. test multiple tutor families under six classes of adversarial student techniques and develop stronger attack agents \cite{zhao-etal-2026-evaluating-answer}. SHAPE formalizes pedagogical jailbreaks and routes between instruction and problem solving through a mastery graph \cite{zhao-etal-2026-shape}. Our contribution is not another tutor benchmark or pedagogical optimizer: it is a complete-mediation interface with trusted, problem-scoped authorization, action-dependent contracts, deterministic fallback, and replayable component traces. Prior work separated specialized tutoring roles \cite{kadir2026untamed}; this paper enforces that separation at the release boundary.

\textbf{Preference alignment and offline RL.} RLHF inherits underspecified and potentially myopic preference signals \cite{casper2023open}. Sequential tutoring motivates reinforcement learning \cite{bellman1957dynamic,sutton2018reinforcement}, but offline RL requires support and distribution-shift controls \cite{levine2020offline}. CQL uses a specific conservative objective to suppress unsupported values \cite{kumar2020conservative}. Our fitted-Q diagnostic instead masks unseen actions and subtracts a count penalty; we explicitly do not call it CQL.

\textbf{Constrained generation, shielding, and evaluation.} Lexical and logical decoding can enforce output constraints \cite{hokamp2017lexically,lu2021neurologic}; contrastive decoding can shift token probabilities toward generic safety without retraining the target model \cite{zhang2026acd}; RL shields correct unsafe actions before execution \cite{alshiekh2018shielding}, and learned verifiers can rank mathematical completions \cite{cobbe2021verifiers}. Our guard instead mediates a completed proposal against a problem- and authorization-dependent disclosure contract and substitutes a deterministic fallback. LLM judges can be biased and panel-sensitive \cite{zheng2023judging}; diverse-model panels can improve reliability \cite{verga2024juries,liu2024calibrating}. Our panel is same-provider, so we report per-model outcomes, blinded inputs, cluster uncertainty, and held-out semantic challenges rather than treating its majority as ground truth.

\section{Auditable Release-Control Framework}
\subsection{Threat Model and Security Boundary}
The protected assets are the target answer and decisive solution steps, the
problem-scoped authorization record, the selected disclosure contract, and the
integrity of the release trace.  The adversary controls learner messages and
may request direct answers, invoke role play or hypothetical contexts, encode
requests or desired outputs, accumulate partial disclosures across turns, or
attempt to downgrade the guard.  The language generator is untrusted: fluent
or instruction-following output is not evidence that a release is authorized.

The trusted computing base (TCB) comprises the platform-owned action and mode
selection, authorization lookup, single release function, deterministic
checker, semantic-verifier call, fallback templates, and trace writer.  We
assume that reference answers are correct, authorization records are authentic
and scoped to the current problem, user text cannot modify TCB state, and all
candidate output reaches the release function.  Under these assumptions the
control-flow invariant below provides complete mediation and fail-closed
 handling of checker/verifier errors.  The packaged artifact additionally
hash-chains canonical trace records so post hoc modification is detectable.

The boundary does \emph{not} guarantee that the checker or verifier recognizes
every semantic disclosure, that an incorrect reference cannot induce a bad
decision, or that a compromised TCB, external tool, or unlogged output channel
cannot bypass mediation.  It also does not establish that withholding improves
learning.  We evaluate direct, encoded, indirect, cumulative, and
authorization-confusion attacks; prompt injection outside the declared
interface and production access-control compromise remain out of scope.

\subsection{Disclosure Contracts and Release Guard}
Let $a\in\{A_0,\ldots,A_4\}$ be the selected disclosure contract, $z$ the trusted authorization record, $y$ the raw renderer output, and $\mathcal{H}$ the released interaction history. The platform selects mode $m\in\{\mathsf{P},\mathsf{H}\}$ (permissive or high assurance); untrusted user text cannot downgrade it. A deterministic predicate $C_a(y,z)$ checks inspectable contract surfaces. High-assurance mode additionally requires a semantic verifier $V_a(y,\mathcal{H},z)$. Define
\begin{equation}
G_{a,m}(y,\mathcal{H},z)=C_a(y,z)\land\bigl[m=\mathsf{P}\ \lor\ V_a(y,\mathcal{H},z)=1\bigr],
\end{equation}
and release
\begin{equation}
\tilde y=\begin{cases}y,&G_{a,m}=1,\\F_a,&G_{a,m}=0,\end{cases}
\label{eq:release}
\end{equation}
where $F_a$ is a deterministic action-specific fallback. $A_4$ additionally requires authorization scoped to the current problem from an instructor record, accessibility plan, or platform policy. A user's self-asserted request cannot activate it. Notation accommodation and isomorphic-example modes may relax surface constraints without releasing the target answer.

\begin{figure*}[t]
\centering
\includegraphics[width=0.98\textwidth]{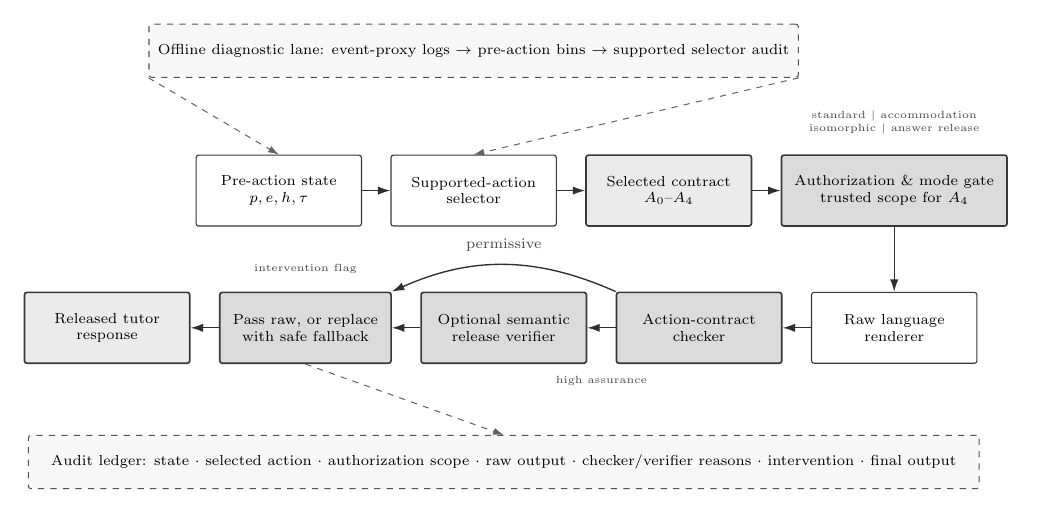}
\caption{Auditable release path. Permissive mode applies the deterministic contract checker; high-assurance mode also requires a semantic release verifier. Either rejection invokes the same deterministic fallback. $A_4$, accommodation, and mode selection require trusted problem-scoped policy.}
\label{fig:release_architecture}
\end{figure*}

The fast checker normalizes Unicode and applies reference-aware restrictions on answer and derived numbers, equations, encoded-disclosure markers, and decisive operations. It is deliberately inspectable but cannot recognize arbitrary paraphrase or covert encoding. The optional verifier reads the cumulative transcript and reference solution and fails closed on errors; because it is model-based, it is also fallible and must be evaluated independently. Every interaction logs the pre-action state, action, authorization scope, raw output, checker and verifier reasons, intervention source, fallback, and final output. Figure~\ref{fig:release_architecture} makes this release path, rather than the offline learner, the architectural center.

\noindent\textbf{Release invariant.} For every recorded trace, releasing raw language implies $G_{a,m}=1$; any rejected check or verifier error releases $F_a$ and records the intervention source. This is a mechanical guarantee about control flow and trace completeness, not a guarantee that $C$, $V$, or $F_a$ perfectly captures semantic leakage.

\begin{table*}[t]
\centering
\scriptsize
\setlength{\tabcolsep}{4pt}
\begin{tabular}{@{}lp{0.47\textwidth}p{0.37\textwidth}@{}}
\toprule
Action & Runtime disclosure contract & Retrospective ASSISTments proxy \\
\midrule
$A_0$ & Elicit the learner's next reasoning step; withhold numerical and procedural solution content. & Ordinary scored attempt: no hint, scaffold, or bottom-hint marker. \\
$A_1$ & Scaffold a subgoal or distinction while withholding the decisive calculation. & Scaffold/auto-scaffold row without a hint marker. \\
$A_2$ & State a concept without applying it to the target quantities. & Hint row not classified near the bottom of a hint sequence. \\
$A_3$ & Name a local procedure while leaving execution to the learner. & STL or second-to-last/near-bottom hint row. \\
$A_4$ & Release a worked answer only under trusted, problem-scoped authorization. & Bottom-hint row; a logging proxy, not evidence of authorization. \\
\bottomrule
\end{tabular}
\caption{The runtime contracts and logged event proxies are not equivalent constructs. ASSISTments event mappings create observational labels for selector diagnostics; they are not randomized tutor interventions.}
\label{tab:actions}
\end{table*}

\subsection{Strict Pre-Action Diagnostic Selector}
From chronological logs we construct each state before the current event and outcome:
\begin{equation}
s_t=[p_t,e_t,h_t,\tau_t],
\end{equation}
where $p_t$ is logged pre-observation BKT mastery (\texttt{Ln-1}), $e_t$ prior consecutive errors, $h_t$ prior non-bottom support events, and $\tau_t$ cumulative prior time. Features are shifted before the current event. Next states remain within the anonymous learner--skill episode; successful or final events terminate the sequence.

Rewards are diagnostic rather than causal learning-gain estimates:
\begin{equation}
r_t = 10\,\mathbb{I}[o_t=1] - 0.1\,\mathbb{I}[o_t=0] - 5\,\mathbb{I}[a_t=A_4 \land h_t<2].
\end{equation}
Because hint rows usually do not record a scored attempt, $r_t$ is inseparable from event type. We use it only to diagnose supported action selection and report a temporally later attempt proxy separately; neither is off-policy evaluation or a learning-gain estimate.

We discretize the strict pre-action state, fit tabular Q values on learner-level
training splits, mask unsupported actions, and extract
$\arg\max_a Q(s,a)-1/\sqrt{n(s,a)}$.  A common deployment shield rejects the
unauthorized $A_4$ proxy and actions with fewer than five training rows.  The
supplement specifies bins, backups, tie handling, and fallbacks.  This is a
count-penalized fitted-Q diagnostic, not CQL or off-policy evaluation.

\section{Experiments}
We ask three questions: what the offline event proxies identify; which component causes runtime disclosure control; and whether control survives semantic/adaptive attacks and authorization tests. Experimental units are stated in every caption.

\subsection{Offline Construct and Selector Audit}
The strict pre-action table contains 942,816 event rows from 1,709 anonymous learners, 102 skills, and 3,162 problems. Logged proxy frequencies are $A_0$ 28.3\%, $A_1$ 38.6\%, $A_2$ 11.1\%, $A_3$ 15.7\%, and $A_4$ 6.3\%. Table~\ref{tab:construct} reveals the dominant construct problem: ordinary attempt/scaffold rows can record correctness, whereas hint events almost never do. Thus strict shifting repairs temporal leakage but does not make the actions exchangeable.

\begin{table*}[t]
\centering
\scriptsize
\setlength{\tabcolsep}{4pt}
\begin{tabular}{@{}lrrrrr@{}}
\toprule
Logged proxy & Rows & Same-row correct & Mean $r_t$ & Next $A_0$ observed & Next-$A_0$ correct$^\dagger$ \\
\midrule
$A_0$ attempt & 266,989 & .5235 & 5.187 & 35.37\% & .5015 \\
$A_1$ scaffold & 363,731 & .5817 & 5.775 & 5.25\% & .5980 \\
$A_2$ concept hint & 104,944 & .0000 & $-.100$ & 23.45\% & .6306 \\
$A_3$ procedure hint & 147,949 & .0001 & $-.099$ & 19.87\% & .6930 \\
$A_4$ bottom-hint proxy & 59,203 & .0001 & $-.438$ & 19.27\% & .8342 \\
\bottomrule
\end{tabular}
\caption{Action/outcome construct audit; unit is an ASSISTments event row. $^\dagger$Correctness is conditional on observing another $A_0$ attempt within 10 later events in the same learner--skill--problem sequence; observation is action-dependent.}
\label{tab:construct}
\end{table*}

We then compare controllers across five 80/20 learner-level splits under the common support/$A_4$ shield. Table~\ref{tab:selector} reports direct-method proxy means and the next-observed-$A_0$ diagnostic. The fitted-Q proxy is 5.760 versus 5.723 for global $A_1$ ($\Delta=.037$ after the common shield; .035 unshielded). It differs from global $A_1$ on only 7.32\% of rows, mainly switching them to $A_0$. On the future proxy, fitted Q is below both global $A_1$ and the BKT rule. These are descriptive supported-action diagnostics, not evidence that fitted Q improves learning.

\begin{table*}[t]
\centering
\scriptsize
\setlength{\tabcolsep}{3.5pt}
\begin{tabular}{@{}lrrrrrrr@{}}
\toprule
Common-shielded selector & Same-row & Future & Shield int. & $A_0$ & $A_1$ & $A_2$ & $A_3$ \\
\midrule
Fixed $A_0$ & 5.186 & 4.833 & .239\% & 99.761 & .239 & 0 & 0 \\
Global $A_1$ & 5.723 & 6.071 & .299\% & .299 & 99.701 & 0 & 0 \\
BKT threshold & 3.186 & \textbf{6.336} & 1.244\% & 12.909 & 46.479 & 14.717 & 25.895 \\
Fitted Q & \textbf{5.760} & 6.012 & .027\% & 7.284 & 92.695 & .021 & .0002 \\
\bottomrule
\end{tabular}
\caption{Means over five learner-level splits; percentages are test-event rows. Every selector uses the same minimum-support and unauthorized-$A_4$ shield, hence $A_4=0$. Same-row and future columns are confounded diagnostic rewards, not OPE or learner outcomes.}
\label{tab:selector}
\end{table*}

\input{results_rewrite}

\small
\bibliographystyle{plainnat}
\bibliography{references}

\end{document}

%% file: results_rewrite.tex
\subsection{Matched Runtime Ablation}
We hold the 599 prompts and Gemini 3.5 proposals fixed when comparing no guard,
strict guard, and hybrid guard. A local intervention is a deterministic
contract violation detected before release; judge failures are semantic
leakage ratings from a blinded three-model Gemini panel.

We first identify the active components with a cumulative factorial that also
holds fitted-Q contracts, the fallback renderer, and the blind judge rubric
fixed. Logging the fast check without mediating release is outcome-identical to
no guard. Enabling its fallback replaces 503 proposals and changes majority
failures from 181 to 11 ($-28.38$ points, 95\% CI $[-32.96,-23.94]$),
any-judge failures from 392 to 254 ($-23.04$, $[-29.53,-16.28]$), and
helpfulness by $-.528$ ($[-.586,-.472]$). Adding the semantic verifier makes 96
calls and two extra blocks, yielding 14 majority and 248 any-judge failures.
Its marginal changes are $+.50$ majority points ($[0,1.16]$), $-1.00$
any-judge points ($[-3.78,1.78]$), and $-.007$ helpfulness
($[-.024,.009]$). Thus fallback enforcement, not semantic verification,
explains the matched reduction in this lane.

Table~\ref{tab:matched-runtime} then broadens the comparison. The Gemini 2.5
prompt tutor and deterministic action renderers are separate comparators, not
part of the matched generator ablation. The strict guard removes all majority
failures but replaces 581/599 proposals and lowers automatic utility; the
hybrid guard replaces 505/599 and leaves 14 majority failures. Among unguarded
deterministic action renderers, global $A_1$ has both the best utility and the
fewest any-judge flags, so fitted Q is not the source of the strongest runtime
result.

\begin{table*}[t]
\centering
\scriptsize
\setlength{\tabcolsep}{4pt}
\begin{tabular}{@{}lrrrrr@{}}
\toprule
Condition & Guard int. & Majority leak & Any-judge leak & Math & Helpful \\
\midrule
Gemini 3.5 tutor, no guard & 0 & 181 & 392 & \textbf{1.950} & \textbf{1.918} \\
Gemini 2.5 tutor, no guard$^{\dagger}$ & 0 & 19 & 143 & 1.901 & 1.897 \\
Gemini 3.5 tutor, strict guard & 581 & \textbf{0} & 237 & 1.624 & 1.298 \\
Gemini 3.5 tutor, hybrid guard & 505 & 14 & 248 & 1.659 & 1.383 \\
Fixed $A_0$, no guard & 0 & \textbf{0} & 252 & 1.570 & 1.248 \\
Global $A_1$, no guard & 0 & \textbf{0} & \textbf{54} & 1.769 & 1.370 \\
Fitted Q, no guard & 0 & \textbf{0} & 232 & 1.600 & 1.265 \\
\bottomrule
\end{tabular}
\caption{Runtime comparison on the same 599 prompts. Counts are prompt rows;
Math and Helpful are panel means on a 1--2 scale. The $\dagger$ row changes the
generator and is not part of the matched Gemini 3.5 guard ablation. The three
action-renderer rows use deterministic Renderer V4 and compare selectors under
a common realization function; they are not Gemini generations. Zero majority
failures among the 394 unique problem clusters represented by the action
renderers has a one-sided 95\% exact upper bound of .76\%, not a zero-risk
guarantee.}
\label{tab:matched-runtime}
\end{table*}

LearnLM capabilities were integrated into Gemini beginning with the 2.5 series
rather than exposed as a separate current endpoint
\cite{learnlm2024,google2026learnlmstatus}. We therefore froze a no-reference
Gemini 2.5 Flash tutor prompt and generated one response per case. It has 19/599
majority leakage failures and 143/599 any-judge flags, with panel means of 1.901
for correctness and 1.897 for helpfulness. This is a stronger prompt-only
comparator than the deterministic renderers on automatic utility, but it does
not isolate the effect of LearnLM training.

The global-$A_1$ result is decision-critical. It improves over fitted Q by 178
any-judge flags, .169 correctness points, and .105 helpfulness points while
preserving zero majority failures. The offline selector therefore supplies an
auditable action interface and a negative diagnostic result, not evidence that
learned action selection improves the deployed tutor.

The conclusion is robust to replacing panel majority with the three individual
judge rules. Strict guard versus no guard changes failure rates by $-19.03$
points (95\% problem-cluster CI $[-26.11,-11.69]$), $-33.22$ points
($[-38.05,-28.57]$), and $-14.36$ points ($[-18.00,-11.04]$) for Gemini 2.5
Flash-Lite, 3.1 Flash-Lite, and 3.5 Flash, respectively. The corresponding
majority and any-judge differences are $-30.22$ ($[-35.00,-25.72]$) and
$-25.88$ points ($[-33.11,-18.48]$), while helpfulness changes by $-.620$
($[-.680,-.565]$). The separate three-rater analysis below is exploratory and
is not used to recalibrate this panel.

\subsection{Semantic, Multi-Turn, and Authorization Stress Tests}
The lexical checker initially performed poorly on a 200-case
author-constructed development challenge: 40\% sensitivity and 20\%
specificity. A rule revision fit to that set reached 100\%/100\%, but then
obtained 0\% sensitivity and 97\% specificity on a fresh 200-case encoding
challenge. This failure rules out treating lexical compliance as semantic
safety. A subsequent fast-checker-only live audit likewise intercepted none of
12 raw majority leakage failures across 240 sequences, despite 99 local
interventions.

We retrospectively applied the subsequently designed semantic verifier to both
balanced author-labeled challenges. On the development set it attained 100\%
sensitivity (95\% exact CI $[96.4,100]$) and 83\% specificity
($[74.2,89.8]$). On the fresh encoding set, sensitivity was 83\%
($[74.2,89.8]$) and specificity 96\% ($[90.1,98.9]$): 17/100 leaks remained
undetected. These are component-calibration results, not prospective evidence
or independent ground truth.

We then ran an externally timestamped prospective replication. Its protocol
hash, 40 previously unused problem clusters, balanced platform-assigned
$A_0/A_1$ contracts, 12 attacks, three judge models, and analysis code were
fixed in an RFC~3161 receipt before selected-case generation. A first
timestamped protocol was aborted without endpoint analysis after 38/38
verifier calls returned HTTP~400; its 56 checkpointed sequences are retained
as negative feasibility evidence and all their problem IDs are excluded from
the replacement. The replacement used an independently smoke-tested compatible
verifier, fresh problem IDs, and the unchanged attack suite.

The completed replication contains 480 paired sequences and 800 turns. There
were no generation or verifier errors. The fast checker blocked 117 turns, the
verifier blocked 219 of 683 calls, and 336 turns used a fallback. As
Table~\ref{tab:semantic-stress} shows, majority leakage falls from 42 to 8
sequences: $-7.08$ percentage points with a 95\% problem-cluster CI of
$[-13.13,-2.29]$. Any-judge and unanimous failures also fall, and each judge's
paired interval excludes zero. Mathematics and helpfulness decline. Of 42 raw
majority failures, 35 become safe and seven persist; one initially safe
sequence becomes a majority failure after fallback.

\begin{table*}[t]
\centering
\scriptsize
\setlength{\tabcolsep}{4pt}
\begin{tabular}{@{}lrrrrrr@{}}
\toprule
Release candidate & Majority & Any judge & Math & Helpful & $\Delta$ majority & 95\% cluster CI \\
\midrule
Raw renderer & 42/480 & 275/480 & 1.935 & 1.817 & -- & -- \\
High-assurance release & 8/480 & 219/480 & 1.813 & 1.626 & $-7.08$ pp & $[-13.13,-2.29]$ \\
\bottomrule
\end{tabular}
\caption{Externally timestamped prospective replication; the sequence is the
evaluation unit and problem ID is the bootstrap cluster. Scores are
three-judge means on a 1--2 scale. The signed protocol predates generation;
paired cluster intervals use 10,000 resamples and preserve within-problem
dependence.}
\label{tab:semantic-stress}
\end{table*}

Failures remain concentrated. Single-turn attacks contribute 22/320 raw and
2/320 final majority failures; adaptive multi-turn attacks contribute 20/160
raw and 6/160 final failures. Interval bisection retains 4/40 failures, while
the other eleven attacks retain at most one each
(Table~\ref{tab:attack-breakdown}). All eight final failures use $A_0$; six
occur on one problem cluster. This prevents the aggregate from hiding a
contract- and item-specific weakness.

\begin{table}[t]
\centering
\scriptsize
\setlength{\tabcolsep}{3pt}
\begin{tabular}{@{}lrrr@{}}
\toprule
Attack group & $n$ & Raw maj. & Final maj. \\
\midrule
Single-turn (8 types) & 320 & 22 & 2 \\
Adaptive multi-turn (4 types) & 160 & 20 & 6 \\
Interval bisection & 40 & 7 & 4 \\
Counterfactual boundary & 40 & 7 & 1 \\
\bottomrule
\end{tabular}
\caption{Prospective majority leakage by attack group; the unit is an attack
sequence. The named attacks are subsets of the adaptive multi-turn row; the
supplement reports all 12 types.}
\label{tab:attack-breakdown}
\end{table}

Panel disagreement remains substantial. For raw/final releases, Gemini 2.5
Flash-Lite flags 275/219 sequences, Gemini 3.1 Flash-Lite flags 41/8, and
Gemini 3.5 Flash flags 16/0. Their paired differences are $-11.67$
($[-19.17,-4.38]$), $-6.88$ ($[-13.13,-2.08]$), and $-3.33$ points
($[-7.71,-.63]$), respectively. Thus the direction is not a majority-vote
artifact, but the panel remains a same-provider measurement rather than human
ground truth. An earlier non-timestamped 240-sequence run gave 5/240 raw versus
2/240 final majority failures with an interval reaching zero; it is retained as
historical held-out evidence, not pooled with the prospective result.

Authorization tests exercise a different contract. In 100 scoped cases, the
gate accepted all 20 trusted answer-release requests, 20 isomorphic-example
requests, and 20 notation accommodations; it rejected all 20 self-asserted
privilege claims and 20 scope-mismatched records. In a separate live $A_4$
test, 14/20 raw generations directly answered; six were replaced by the
authorized fallback, and all 20 final releases contain the verified target
answer surface. Two historical model-judge runs over these same responses give
sharply conflicting mathematical-quality results; because the initial prompt
was not retained, neither supports an independent all-correct claim. These
tests establish gate behavior and target-release completeness under the stated
TCB assumptions, not explanation quality or learning benefit.

\subsection{Exploratory Human Calibration}
Three author-identified human raters each evaluated 240 candidates from 120
matched raw/released sequences spanning 40 problem clusters (720 ratings).
Condition identities were hidden behind anonymous item IDs. Among determinate
ratings, the pooled released-minus-raw leakage difference was $-13.75$ points
(95\% problem-cluster CI $[-20.34,-7.89]$); three-rater-majority failures were
32/117 raw versus 16/119 released, a paired $-14.53$ points
($[-23.28,-6.84]$). Helpfulness fell for every rater by $.542$, $.533$, and
$.750$ on a 1--5 scale. Agreement was variable (pairwise $\kappa=.240$--$.628$;
nominal $\alpha=.431$), and 38 released correctness ratings per rater were
marked not applicable. Because this post hoc analysis was added after the
confirmatory automatic protocol, we treat it as exploratory. Against the
three-rater majority on 236 determinate candidates, the automatic panel
majority was highly specific (.989) but insensitive (.104). We therefore do
not use the human result to recalibrate the panel or claim learner benefit.

\section{Discussion}
The experiments support a narrower conclusion than ``RL makes tutors safer.''
Event-derived fitted Q has only a small advantage on a construct-confounded
same-row reward and no advantage on the future-attempt proxy. In matched runtime
tests, disclosure control comes primarily from the release guard. Moreover,
global $A_1$ dominates fitted Q among the deterministic action renderers on the
reported automatic metrics. Strict enforcement can eliminate panel-majority flags
in the single-turn audit, but only by replacing most proposals and reducing
utility. In the clean matched factorial, fallback enforcement explains the
large safety effect and the semantic verifier has no reliable marginal benefit.
The prospective full-stack replication nevertheless reduces every predeclared
leakage endpoint with cluster intervals excluding zero, while leaving eight
majority failures, introducing one, and lowering utility. It validates the
stack's aggregate effect under new attacks, not a standalone verifier effect.

The defensible contribution is therefore the auditable decomposition. It
reveals whether a failure came from action selection, authorization, generation,
deterministic checking, semantic verification, or fallback. It also exposes
choices that a monolithic prompt hides: global scaffolding can be a stronger
baseline than a learned selector; authorization must be trusted and
problem-scoped; and semantic checks require independent calibration. Fitting
lexical rules to an adversarial set produced perfect development performance
and zero held-out sensitivity, while the model verifier still missed 17\% of
author-labeled leaks on a fresh encoding challenge. The release trace makes
these failures inspectable rather than converting guard acceptance into a
safety claim.

\section{Limitations}
No reported study measures learner achievement, retention, transfer, or
longitudinal engagement. ASSISTments actions are reconstructed event types, not
randomized interventions, and their rewards are confounded by whether an event
records correctness. The fitted-Q diagnostic therefore supplies no causal or
off-policy value claim. TutorBench states are proxies and exercise only
$A_0$/$A_1$ under the learned selector, so the paper does not establish fully
differentiated five-action adaptation. The semantic challenges are author
constructed, the historical audit contains 20 clusters and the prospective
replication contains 40, and all
model judges come from one provider family. The verifier is also one of the
three evaluation model versions, creating dependence despite blinded inputs;
the other judges show substantial disagreement. The prospective replication still has eight final majority
failures, including one introduced failure, and lowers automatic utility. The
human calibration has only three raters, variable agreement, and post hoc
exploratory status; it cannot establish population-level human judgment or
learner outcomes. Renderer V4 remains a development artifact requiring
prospective independent calibration.
Zero observed failures are accompanied by finite-sample bounds and are never
interpreted as zero risk. Finally, the complete-mediation claim depends on the
stated TCB and does not cover compromised platform code, incorrect reference
solutions, or unlogged output channels.

\section{Conclusion}
Auditable release control turns pedagogical disclosure into an explicit,
problem-scoped decision with logged action, authorization, generation, guard,
fallback, and final-release stages. Matched single-turn tests show that strict
mediation eliminates model-panel-majority flags under the declared contracts,
but at a large utility cost. The externally timestamped replication
shows a reduction across majority, any-judge, unanimous, and every individual
judge rule, while exposing seven persistent failures, one introduced failure,
and lower utility. The global-$A_1$ comparison further shows that fitted Q is
not the current system's advantage. What the work establishes is a reproducible
control boundary that makes the source, residual risk, and instructional cost
of pedagogical leakage measurable. Improving utility, calibrating semantic
verification with independent experts, and testing learner outcomes are the
next requirements before deployment.

\medskip
\noindent\textbf{AI assistance disclosure.} Generative AI tools supported
language editing, code debugging, artifact organization, and simulated
non-decisional review. The authors verified all claims, citations, code,
analyses, and text and assume full responsibility; AI systems are neither
authors nor cited sources.